\documentclass{lhep}    
\usepackage[colorlinks,linkcolor=blue,citecolor=blue,urlcolor=blue,bookmarks,bookmarksnumbered]{hyperref}

\usepackage{amsfonts,mathrsfs,amssymb}
\usepackage{amsmath,amsthm,bm,multirow,longtable}

\newcommand{\be}{\begin{equation}}
\newcommand{\ee}{\end{equation}}
\newcommand{\bea}{\begin{eqnarray}}
\newcommand{\eea}{\end{eqnarray}}
\newcommand{\ba}{\begin{array}}
\newcommand{\ea}{\end{array}}
\newcommand{\ben}{\begin{enumerate}}
\newcommand{\een}{\end{enumerate}}
\newcommand{\bit}{\begin{itemize}}
\newcommand{\eit}{\end{itemize}}
\newcommand{\bc}{\begin{center}}
\newcommand{\bfig}{\begin{figure}}
\newcommand{\efig}{\end{figure}}
\newcommand{\bq}{\begin{quotation}}
\newcommand{\eq}{\end{quotation}}
\newcommand{\bt}{\begin{table}}
\newcommand{\et}{\end{table}}
\newcommand{\btab}{\begin{tabular}}
\newcommand{\etab}{\end{tabular}}
\newcommand{\bs}{\begin{slide}}
\newcommand{\es}{\end{slide}}

\newcommand{\pa}{\partial}

\newcommand{\IR}{\mathbb{R}}

\newcommand{\X}{\mathbb{X}}

\newcommand{\rd}{\mathrm{d}}
\newcommand{\Tr}{\mathrm{Tr}}

\def\S{\mathbb{S}}
\def\s{\sigma}

\def\S{\mathbb{S}}

\def\S{\mathbb{S}}

\let\a=\alpha
\def\tt{\tilde{\tau}}
\def\ts{\tilde{\sigma}}

\let\ba=\overline

\def\const{\textit{const.\,}}
\def\rd{{\rm d}}
\def\define{\buildrel{\smash{\scriptscriptstyle\rm def}}\over=}

\let\F=\Phi
\def\e{\epsilon}
\let\F=\Phi
\let\g=\gamma
\def\inv#1{{\textstyle{1\over#1}}}

\let\j=\psi

\let\l=\lambda
\let\L=\Lambda

\let\q=\theta

\let\p=\pi
\let\t=\tau
\let\vd=\partial

\def\hx{\mathord{\hat x}}
\def\tx{\mathord{\tilde x}}
\def\htx{\mathord{\hat{\tilde x}}}
\let\z=\zeta

\let\w=\omega

\def\frc#1#2{{\textstyle{#1\over#2}}}
\def\IR{\relax\leavevmode{\rm I\kern-.18em R}}
\def\ZZ{\relax\leavevmode
       \ifmmode\mathchoice
       {\hbox{\sf Z\kern-.4em Z}}
       {\hbox{\sf Z\kern-.4em Z}}
       {\lower.9pt\hbox{\scriptsize\sf Z\kern-.36em Z}}
       {\lower1.2pt\hbox{\tiny\sf Z\kern-.36em Z}}
       \else{\sf Z\kern-.4em Z}\fi}
\def\RR{\relax\leavevmode
       \ifmmode\mathchoice
       {\hbox{\sf R\kern-.4em R}}
       {\hbox{\sf R\kern-.4em R}}
       {\lower.9pt\hbox{\scriptsize\sf R\kern-.36em R}}
       {\lower1.2pt\hbox{\tiny\sf R\kern-.36em R}}
       \else{\sf R\kern-.4em R}\fi}

\def\resetby#1#2{\@addtoreset{#2}{#1}}
\def\seceq{\@addtoreset{equation}{section}
              \def\theequation{\thesection.\arabic{equation}}}

\def\Label#1{\label{#1}%
                \smash{\hbox to0pt{\raise1ex\hbox{\tiny[#1]}\hss}}}
\def\noLabels{\let\Label=\label}

\def\TeV{\text{T\kern0pte\kern-1ptV}}

\def\brb#1#2{[\![\mkern2mu#1,\mkern-2mu#2\mkern2mu]\!]}

\parskip\medskipamount

\journal{LHEP}
\vol{xx}
\jyear{2021}
\pages{xxx} 

\received{xx October 2020}
\published{xx 2021}

\def\be{\begin{equation}}
\def\ee{\end{equation}}
\def\bea{\begin{eqnarray}}
\def\eea{\end{eqnarray}}

\begin{document}

\title{String Theory, the Dark Sector 
and the Hierarchy Problem}

\author{Per Berglund,\auno{1} Tristan H{\"u}bsch,\auno{2} and Djordje Minic\auno{3}}
\address{$^1$Department of Physics and Astronomy, University of New Hampshire,
 Durham, NH 03824, USA}
\address{$^2$Department of Physics and Astronomy, Howard University, Washington,
 D.C., 20059, USA}
\address{$^3$Department  of Physics, Virginia Tech, Blacksburg, VA 24061, USA}

\begin{abstract}
We discuss dark energy, dark matter and the hierarchy problem in the context of a general non-commutative formulation of string theory. In this framework 
dark energy is generated by the dynamical geometry of the dual
spacetime while dark matter, on the other hand, comes from the degrees of freedom dual to the visible matter.
This formulation of string theory is sensitive both to the IR and UV scales and the Higgs scale is radiatively stable by being a geometric 
mean of radiatively stable UV and IR scales.
We also comment on various phenomenological signatures of this novel approach to dark energy,
dark matter and the hierarchy problem. 
We find that this new view on the
hierarchy problem is realized in a toy model based on a non-holomorphic deformation of the stringy cosmic string.
Finally, we discuss a proposal for a new
non-perturbative formulation of string theory, which sheds light on M theory and F theory, as well as on supersymmetry and holography.
\end{abstract}

\maketitle

\begin{keyword}
string theory\sep dark energy\sep dark matter\sep hierarchy problem
\doi{}
\end{keyword}

\section{Introduction}

The deep foundations and real world implications of string theory~\cite{Polchinski:1998rq} are still shrouded in mystery.
In particular, the problem of dark energy has excited a lot of recent discussion within various aspects of string theory
\cite{Danielsson:2018ztv, Obied:2018sgi, Andriot:2019wrs}.
Motivated in part by some prescient work on various aspects of string theory
\cite{Casher:1985ra, Atick:1988si, Tseytlin:1990nb, Moore:1993zc},
as well as some recent developments in non-commutative field theory~\cite{Douglas:2001ba}, 
double field theory~\cite{Hull:2009mi}, quantum gravity~\cite{AmelinoCamelia:2011bm}
and quantum foundations~\cite{aharonov2008quantum, Minic:2003nx},
a generic, non-commutatively generalized geometric phase-space formulation of string theory~\cite{Freidel:2013zga}
has been recently developed.
Within this framework, which clarifies many foundational issues found in the
classic textbook discussion of string theory~\cite{Polchinski:1998rq}, it was recently shown that dark energy is naturally induced from the overall curvature of the dual of the observed spacetime~\cite{Berglund:2019ctg}.
This effect is the leading (zeroth) order term in an expansion of the non-commutative scale, $\lambda$, and is realized in certain stringy-cosmic-string-like toy models that rely only on the ubiquitous axion-dilaton (``axilaton'') system and gravity~\cite{rBHM7}. 
In particular, it was pointed out that the Higgs scale is radiatively stable by being the geometric mean of the radiatively stable UV (Planck) and IR (dark energy) scales, respectively.

In this paper we consider the inclusion of the dual matter which, to first order in $\lambda$, can be interpreted as the dark matter as well as providing an intrinsic stringy resolution to the hierarchy problem.
Furthermore in this new approach, the known Standard Model fields and their dark matter duals are dynamically correlated, and we comment on various phenomenological signatures stemming from this correlation.
We also comment about other phenomenological signatures of this new approach to dark energy,
dark matter and the hierarchy problem in the context of string theory. 
In particular, we discuss this new view on the
hierarchy problem within a toy model based on a non-holomorphic deformation of the stringy cosmic string.
Finally, we present a proposal for a new
non-perturbative formulation of string theory, which sheds light on M theory as well as F theory, and illuminates
the emergence of supersymmetry and holography.

\section{General string theory and dark energy} 

We begin our discussion with a review of the generic non-commutative and doubled formulation of string
theory and its relation to dark energy. Following the recent discussion
in~\cite{Freidel:2013zga},
the starting point is the 
{\it chiral} string worldsheet description
\be
S^{\text{ch}}_{\text{str}}\,{=}\,\frac{1}{4\pi}\int_{\t,\s}\!
 \Big[\pa_{\tau}{\X}^{A} (\eta_{AB}+\w_{AB})
- \pa_\s\X^A H_{\!AB}\Big] \pa_\s\X^B, 
\label{e:MSA}
\ee 
with worldsheet coordinates $\t,\s$, and $\X^A(\t,\s)$, ($A=1,\dots,26$, for the critical bosonic string) combine the sum ($x^a$) and the difference ($\tx_a$) of the left- and right-moving {\em\/chiral bosons\/} on the string.\footnote{This formulation also leads to a natural proposal for a non-perturbative string theory which will be discussed
in section~\ref{s:non-pert} of this paper.}
The mutually compatible dynamical fields 
are
the antisymmetric symplectic structure $\w_{AB}$,
the symmetric polarization metric $\eta_{AB}$ and
the doubled symmetric metric $H_{\!AB}$, respectively, defining the so-called Born geometry ~\cite{Freidel:2013zga}.
Quantization renders the doubled ``phase-space'' operators $\hat{\X}^A=(\hx^a/\l, \htx_a/\l)$ inherently non-commutative, inducing~\cite{Freidel:2013zga}
\be
[ \hat{\X}^A, \hat{\X}^B] = i \w^{AB},
\label{e:CnCR}
\ee
or, in components, for constant non-zero $\w^{AB}$,
\bea
[\hx^a,\htx_b]=2\pi i\l^2 \delta^a_b,\quad
[\hx^a,\hx^b]=0=[\htx_a,\htx_b],
\label{e:CnCR1}
\eea
where $\l$ denotes the fundamental length scale, such as the fundamental Planck scale,
$\e=1/\l$ is the corresponding fundamental energy scale and the string tension is 
$\a' = \l/{\e}=\l^2$; see also Sect.~\ref{s:BHM}.
Note that the Hamiltonian and diffeomorphism constraints, $\pa_\s\X^A H_{AB} \pa_\s\X^B =0$
and $\pa_\s\X^A \eta_{AB} \pa_\s\X^B =0$, respectively, are treated on the equal footing.
In particular, the usual spacetime interpretation of the zero mode sector of string theory~\cite{Polchinski:1998rq} is
tied to the solution of the diffeomorphism constraint by level matching. In this more general and generically
non-commutative formulation, the spacetime interpretation is replaced by a modular spacetime (or quantum spacetime)
realization (also found in the context of quantum foundations)~\cite{Freidel:2013zga}.
Thus, all effective fields must be regarded as 
bi-local $\phi(x, \tx)$~\cite{Freidel:2013zga}, subject to~\eqref{e:CnCR1}, and therefore inherently non-local in 
the conventional $x^a$-spacetime. Such non-commutative field theories~\cite{Douglas:2001ba} 
generically display a mixing between the ultraviolet (UV) and infrared (IR) physics
with continuum limits defined via a double-scale renormalization group (RG) and the 
self-dual fixed points~\cite{Douglas:2001ba},~\cite{Freidel:2013zga}. 
This has profound implications for the generic physics of string theory, and in particular the problems
of dark energy, dark matter and the separation of scales that goes beyond the realm of effective field theory.

In ~\cite{Berglund:2019ctg} 
we have argued that the generalized geometric formulation of string theory discussed above provides for an effective description of dark energy that is consistent with a de Sitter spacetime. This is due to the theory's chirally and non-commutatively~\eqref{e:CnCR1} doubled realization of the target space 
and the stringy effective action on the doubled non-commutative~\eqref{e:CnCR1} spacetime $(x^a,\tx_a)$ 
\be
  S_{\text{eff}}^{\textit{nc}}
  =\iint \text{Tr} \sqrt{g(x,\tx)}~ \big[R(x,\tx) +L_m(x,\tx) +\dots\big],
\label{e:ncEH}
\ee
including the matter Lagrangian $L_m$ and with the
corresponding Planck lengths  set to one. 
(The ellipses denote higher-order curvature terms induced by string theory.) This result can
be understood as a generalization of the famous calculation by Friedan~\cite{rF79b}. 
Using~\eqref{e:CnCR1}, $S_{\text{eff}}^{nc}$ clearly expands into numerous terms with different powers of $\l$, which upon $\tx$-integration and from the $x$-space vantage point produce
various effective terms.
Dropping $L_m$ 
for now, to lowest (zeroth) order of the expansion in the non-commutative parameter $\l$
of $S_{\text{eff}}^{\textit{nc}}$ takes the form 
\be
S = - \iint\! \sqrt{-g(x)} \sqrt{-\tilde{g}(\tx)}~\big[R(x) + \tilde{R}(\tx)\big],
\label{e:TsSd}
\ee
a remarkable result which first was obtained almost three decades ago by Tseytlin~\cite{Tseytlin:1990nb},  effectively neglecting $\w_{AB}$ in~\eqref{e:CnCR} 
by assuming that $[\hat x,\htx]=0$~\cite{Tseytlin:1990nb}.

In this leading limit, the $\tx$-integration in the first term of~\eqref{e:TsSd} defines the gravitational constant $G_N$, and in the second term produces a {\it positive} cosmological constant $\L>0$. Hence, the weakness of gravity is determined by the size of the canonically conjugate dual space, while the smallness of the cosmological constant is given by its curvature.  In particular we have:
\be
 \bar{S} 
= \frac{\int_X\! \sqrt{-g(x)}\, \big[R(x) +\dots\,\big]}{ \int_X\! \sqrt{-g(x)}}+\dots
 \label{e:TsLb11}
\ee
which leads to a seesaw like formula for the cosmological constant, discussed below.

So far, we have omitted the matter sector explicitly. In what follows, we argue the dual part of the matter sector 
appears as dark matter, which is in turn both sensitive to dark energy~\cite{Ho:2010ca}
and also dynamically correlated with the
visible matter. We next focus on this dark matter, and emphasize the unity of the description of the entire dark energy and matter sector, induced and determined by the properties of the dual spacetime, as predicted by this general, non-commutatively phase-space doubled formulation of string theory.

\section{Dark sector and the hierarchy problem}

We have already emphasized that in this generic non-commutative formulation of string theory,
all effective fields must be regarded a priori as 
bi-local $\phi(x, \tx)$~\cite{Freidel:2013zga}, subject to~\eqref{e:CnCR1}.
Moreover, the fields are doubled as well, and thus for every  $\phi(x, \tx)$ there exists 
a dual $\tilde{\phi}(x, \tx)$. This can be easily seen in the background field approach, which we will
consider in the next section.
Therefore, in general,
to lowest (zeroth) order of the expansion in the non-commutative parameter $\l$
$S_{\text{eff}}^{\textit{nc}}$ takes the following form (that also includes the matter sector and its dual),
which generalizes equation \eqref{e:TsSd}
(and where, once again, the
corresponding Planck lengths are set to one):
\be
\!\!\!
\iint\!\! \sqrt{{g(x)}\,{\tilde{g}(\tx)}} 
 \Big[R(x) {+} \tilde{R}(\tx){+} L_m (A(x, \tilde{x})) {+} \tilde{L}_{dm}   ( \tilde{A}(x, \tilde{x})) \Big].
\label{e:TsSd1}
\ee
Here the $A$ fields denote the usual Standard Model fields, and the $\tilde{A}$ are their duals, as predicted by the general formulation of 
quantum theory that is sensitive to the minimal length (the non-commutative parameter $\l$~\cite{Freidel:2013zga}).
In the following section we will elaborate more explicitly on the dual matter degrees of freedom.
Right now we are concerned with the generalization of the discussion summarized in the previous section.

After integrating over the dual spacetime, and after taking into account T-duality, equation 
\eqref{e:TsLb11} now reads:\footnote{Tseytlin's proposal has been further explored by Davidson and Rubin who in particular showed that the cosmological constant is necessarily non-negative definite~\cite{Davidson}.}
\be
 \bar{S'} 
= \frac{\int_X\! \sqrt{-g(x)}\, \big[R(x) +L_m(x) + \tilde{L}_{dm} (x)\big]}{ \int_X\! \sqrt{-g(x)}}+\dots
 \label{e:TsLb1}
\ee
The proposal here is that the dual sector (as already indicated in the previous section) should be interpreted as the dark matter sector,
which is correlated to the visible sector via the dark energy sector, as discussed in~\cite{Ho:2010ca}.
We emphasize the unity of the description of the entire dark sector based on 
the properties of the dual spacetime, as predicted by the generic formulation of string theory (as a quantum theory with a dynamical Born geometry)~\cite{Freidel:2013zga}.

Let us turn off the dynamical part of gravity and consider the hierarchy problem.
First we have
\be
S_0= - \iint\! \sqrt{{g(x)}{\tilde{g}(\tx)}}~\big[ L_m (A(x, \tilde{x})) + \tilde{L}_{dm}   ( \tilde{A}(x, \tilde{x})) \big],
\label{e:TsSd2}
\ee
which leads to the following non-extensive action to lowest order in $\L$
\be
 \bar{S}_0
= \frac{\int_X\! \sqrt{-g(x)}\, \big[L_m(x) + \tilde{L}_{dm} (x)\big]}{ \int_X\! \sqrt{-g(x)}}+\dots
 \label{e:TsLb2}
\ee
after integrating over the dual spacetime.

This implies a seesaw formula which involves the matter scale from the matter (and dark matter) part of the action
and the scales related to the UV (Planck scale, $M_P$) and the IR (dark energy scale, $M_\L$):
\be
 M_\L   M_P \,{\sim}\,M_H^2 .
\ee
where $M_H$ denotes the characteristic matter scale.
Completely generally, 
this relation follows from the diffeomorphism constraint of the chiral string worlsheet,
which is controlled by the $O(d,d)$ bi-orthogonal metric $\eta$, and which implies, in the limit of zero modes
(and zero momenta)
that $E \tilde{E} = M^2$. Here $E$ and $\tilde{E}$ are the  energy scales in the observed and dual spacetime, respectively, and $M$ is a new (mass) parameter. Since the IR (dark energy) scale can be interpreted as the vacuum energy, which in the matter sector is controlled by the Higgs potential, 
$M$ naturally sets the Higgs scale, $M_H$, which is indeed the case numerically.
If we remember that the geometry of the dual spacetime is responsible for the origin of dark energy, then the
dual energy $\tilde{E}$ can be set to the dark energy scale. Then the fundamental energy scale $E$ in the
observed spacetime is the Planck scale. This mixing of the UV and IR scales in a fully covariant formulation
is a  unique feature of the chiral string worldsheet theory. 
(For other approaches to the hierarchy problem which mix the UV and IR scales but which violate covariance,
consult~\cite{Craig:2019zbn}.)

Note that both the UV and IR scales are radiatively stable.
First, we note that the $M_P$ is the UV scale and the issue of
radiative stability does not apply to it.
Second, let us recall the radiative stability of the dark energy scale $M_\L$, the IR scale:
In particular, as we have noticed in our previous work, the effective action of the sequester type~\cite{Kaloper:2013zca}
(see also~\cite{Kaloper:2014dqa})
\begin{equation}
 \int_x\!\sqrt{-g}~\Big[\frac{R}{2G} + s^4 L (s^{-2} g^{ab}) + \frac{\Lambda}{G}\Big]
 + \sigma\Big(\frac{\Lambda}{s^4 \mu^4}\Big),
\end{equation}
where $L$ denotes the combined Lagrangians for the matter and
dark matter sectors, $\mu$ is a mass scale and $\sigma(\frac{\Lambda}{s^4 \mu^4})$ is a
global interaction that is not integrated over
\cite{Kaloper:2013zca}. 
This can be provided by our set up: Start with bilocal
fields  $\phi(x, \tx)$~\cite{Freidel:2013zga},
and replace the dual labels $\tx$ and also $\l$ (in a coarsest approximation) by the global dynamical scale $s \sim \Delta {\tx} \,{\sim}\,\l^2 \Delta {x}^{-1} $.
Also, normal ordering produces $\sigma$.
This is 
an effective realization of the sequester mechanism 
in a  non-commutative phase of string theory. 
One important lesson here is, that the low energy effective description of the generic string theory is, to lowest order,
a sequestered effective field theory, and more generally, a non-commutative effective field theory~\cite{Douglas:2001ba} of a new kind, which is
defined within a doubled RG, which is covariant with respect to the UV and IR cut-offs, and which is endowed with a
self-dual fixed point~\cite{Freidel:2013zga}.

\section{Dark matter and string theory}

In this section we elaborate on the dual matter degrees of freedom and the explicit appearance of dark matter in
the general formulation of string theory~\cite{Freidel:2013zga}.
The previous discussion looked at the stringy effective action to lowest order in $\L$, neglecting the non-commutative aspect of the generalized geometric description of the string worldsheet ~(\ref{e:MSA}). In order to see the effect of the leading order correction in $\L$ of~(\ref{e:TsSd1})
we consider the zero modes of $S^{\text{ch}}_{\text{str}}$. 
The associated 
particle action is fixed by the symmetries of the chiral worldsheet in terms of the symplectic form $\omega$, the $O(D,D)$ metric $\eta$ and the double metric $H$--the so-called Born geometry--and takes the form (following the general results for the chiral string worldsheet description~\cite{Freidel:2013zga}, 
\be
S_{MP} = \int_\t\!\big( p \dot{x} + \tilde{p} \dot{\tilde{x}} - \l^2 p \dot{\tilde{p}} - N h - \tilde{N} d\big)
\label{e:SMP}
\ee
Here the Hamiltonian constraint, fixed by the double metric, is given by $h = p^2 + {\tilde{p}}^2 + m^2 $, and
the diffeomorphism constraint, fixed by the $O(D,D)$ metric, is $d = p \tilde{p} - M^2$.
These constraints are inherited from the quantization of chiral worlsheet theory. Finally, the symplectic structure fixes
the $ \l^2 p \dot{\tilde{p}} $ term.
Note that $m^2$ should not be confused with the (mass)$^2$ of a particle excitation. In parallel with the usual 
discussion found in introductory chapters of textbooks on quantum field theory 
one has to understand the representation theory associated with the symmetries of the underlying Born geometry,
and interpret $m^2$ and $M^2$ in terms of the relevant Casimirs in the full representation theory of this
description~\cite{fkglm}. 
We are not going to pursue this question in what follows, but we alert the reader that $M$ is a new
parameter not found in the context of the effective field theory description, while $m^2$ can be interpreted
as a particle mass only in a very degenerate limit in which $\tilde{p} =0$ and $M=0$.

For concreteness, let us start with a vector background, by shifting the momenta and the dual momenta by the standard, minimally coupled gauge field and its dual~\cite{Freidel:2013zga}: 
\be
p_{a} \to p_{a} + A_{a}(x, \tilde{x}) , \qquad {\tilde{p}}_{a} \to {\tilde{p}}_{a} + {\tilde{A}}_{a}(x, \tilde{x})
\label{e:minC}
\ee
These gauge fields have the usual Abelian gauge symmetries.
Thus in the target space action for the gauge fields we end up with
canonical Maxwellian terms (with the obvious index structure)
plus a characteristic coupling inherited from the symplectic structure
\be
\int_{x,\tilde{x}} \big[ F^2 -a\l^2\brb{A}{\tilde{A}} +{\tilde{F}}^2+ F \tilde{F} +\dots\big]
\label{e:SMP2}
\ee
and where the $\l^2\brb{A}{\tilde{A}}$ term stems directly from the $\l^2p\dot{\tilde{p}}$-term in~\eqref{e:SMP} by the ``minimal coupling'' shift~\eqref{e:minC}, its dimensionful coefficient $a$ to be determined below.
This ``mixing'' term may be expressed as:
\begin{alignat}9
&\brb{A}{\tilde{A}}\define\int_{\t_*}^\t\rd\t'~
  A(x(\t'))\,\frac{\rd\tilde{A}(\tilde{x}(\t'))}{\rd\t'},\\
&=\frc12\Big[A(x)\tilde{A}(\tilde{x})\Big]_{\t_*}^\t
  +\frc12\int_{\t_*}^\t\rd\t'~
    \Big[A(x)\dot{\tilde{A}}-\dot{A}(x)\tilde{A}(\tilde{x})\Big].
 \label{e:AdotA}
\end{alignat}
The first of these includes a (worldline-local) mixing term, together with its value at some ``reference'' proper time, $\t_*$, the integral of which~\eqref{e:SMP2} evaluates to an irrelevant additive constant. The second part, $[A\dot{\tilde{A}}\,{-}\,\dot{A}\tilde{A}]$, is far more interesting, as it provides a telltale ``Zeemann''-like coupling of $(A,\tilde{A})$-pairs to corresponding ``external/background'' fluxes, scaled by the coefficient ``$a\l^2$.'' This relies on identifying $A,\tilde{A}$ as canonical coordinates in the target-spacetime (classical) field theory. Alternatively, in the underlying worldsheet quantum field theory, the $A,\tilde{A}$ are coefficients of certain quantum states, for which the $\brb{A}{\tilde{A}}$-term likewise accompanies a Berry-phase like quantity.

Second-quantization of the action~\eqref{e:SMP2} in the Coulomb gauge (and its dual) produces the following structure fixed by Born geometry:
\begin{equation}
  \int_{x,\tilde{x}} \big[(\partial A)^2 {+}(\tilde{\partial} {A})^2
 {-}a\l^2\brb{A}{\tilde{A}} {+}(\tilde{\partial} \tilde{A})^2 {+}({\partial} \tilde{A})^2
 {+}\partial A\,\tilde{\partial} \tilde{A} +\dots\big]
\end{equation}
Integrating over the dual spacetime ($\tilde{x}$), and setting $\tilde{p} \to 0$ in the
observable spacetime for consistency, poduces (with indices on the respective gauge fields fully restored)
\be
\int_x \Big[(\vd_{[a} A_{b]})^2  -\frac{\l^8}{L^{10}} f^a_b\brb{A_a}{\tilde{A}^b} 
+({\vd_{[a}} \tilde{A}_{b]})^2 +\dots\Big],
\label{e:SMPv}
\ee
where $\tilde{A}_a\,{\define}\,\eta_{ab}\tilde{A}^b$ and $f^a_b$ encode ``background fluxes,'' naturally at the fundamental scale $\l$ from~\eqref{e:CnCR1}.
 Properly normalized on the world-line~\eqref{e:AdotA}, the ``mixing'' term $\brb{A_a}{\tilde{A}^a}$ in spacetime must be re-normalized by the volume of the primordial observable spacetime, $L^{10}$, prior to any compactification and inflation.
 This term is thereby sensitive both to the {\em\/fundamental\/} UV cutoff $\l$ as well as the {\em\/primordial\/} IR cutoff $L$.

In other words, in the observable spacetime, the visible sector and its dual/dark counterpart
are ``mixed/correlated''~\eqref{e:SMPv}, in a way that is sensitive to both the UV and IR cutoffs. This correlation becomes invisible in effective field theory, and it
vanishes as either $\l \to 0$  or $L \to \infty$. Note that with a particular double scaling,
this term can be finite! This is consistent with the general set-up of the chiral string worldsheet theory
which has two
cutoffs (UV and IR) and which generically should be defined with self-dual RG fixed points.
Such a correlation between visible and dark matter involving an IR scale (in this case, the Hubble scale)
 has been observed in astronomical data and has been studied in the context of modified dark matter
in~\cite{Ho:2010ca}.
However, the ratio $\frac{\l^8}{L^{10}}$ should not be naively considered
to be trans-Planckian, because that would require that $\l$ is the Planck length,
and $L$ is the Hubble length. Instead, the two scales should be considered as effective UV and
IR scales, the joint appearance of  which is a direct evidence of a departure from effective field theory, as further discussed in Sect~\ref{s:BHM}.

Having illustrated the general idea with the vector fields $A_a$ and $\tilde{A}^a$, the
form of the corresponding action for a scalar $\phi$ and its dual $\tilde{\phi}$ is immediate (and similarly for pseudo-scalars):
\be
\int_x \Big[(\partial \phi)^2  - \frac{\l^8}{L^{10}} \brb{\phi}{\tilde{\phi}}
+  ({\partial} \tilde{\phi})^2 +\dots\Big],
\label{e:SMPb}
\ee
as this would be forced in every dimensional reduction framework.
Similarly, the corresponding action for the fermions (by taking the ``square root'' of the propagating part of the scalar action to accommodate 
the spin-statistics theorem)
\be
\int_x \Big[(\bar{\psi} \partial \psi) 
 -\frac{\l^9}{L^{10}} \brb{\bar\psi}{\tilde{\psi}}
 +(\bar{\tilde{\psi}}{\partial} \tilde{\psi}) +\dots\Big],
\label{e:SMPf}
\ee
where $\brb{\bar\j}{\tilde\j}=\bar\j\tilde\j$ are non-derivative bilinear terms, accompanied by ``external'' fluxes as in~\eqref{e:SMPv}. This result may be justified by target-spacetime supersymmetry, even if supersymmetry is ultimately broken: The indicated terms are restricted to free fields in flat-spacetime. In particular, the omitted interaction terms here also include metric and curvature deviations from flat spacetime.
 From the underlying worldline, or even worldsheet~\cite{Freidel:2013zga}) point of view, such terms are induced from generalizing~\eqref{e:SMP} along the standard construction of GLSMs~\cite{rPhases}; the ``free-field-limit'' terms shown herein however remain unchanged.
 The same pre-factor is also implied from the worldline point of view: the superpartner of each $p\dot{\tilde{p}}$-term in the  Lagrangian~\eqref{e:SMP} is a  fermionic bilinear, $\bar\chi\widetilde\chi$, which couples to the same ``external/background'' flux~\eqref{e:SMPb}, giving rise to a worldline super-Zeemann effect~\cite{Doran:2008xw}.

The $\frac{\l^9}{L^{10}}$ scaling coefficient in~\eqref{e:SMPf}, forced on dimensional grounds in this mass-mixing term, sets the scale in this novel ``seesaw mechanism,'' in principle tunable to induce naturally small neutrino masses. In fact, the inclusion of several mass-scales enables concrete models to incorporate an entire hierarchy of seesaw mechanisms, with a more reasonable chance to approach the intricacies of a realistic mass spectrum.

To summarize, the leading ``kinetic'' parts in the actions~\eqref{e:SMPb} and~\eqref{e:SMPf}, together with the ``mixing terms,'' $\brb{\phi}{\tilde\phi}$ and $\brb{\bar\psi}{\tilde\psi}$, are seen to be natural:
 ({\small\bf a})~by dimensional reduction from~\eqref{e:SMPv} to~\eqref{e:SMPb}, and
 ({\small\bf b})~by supersymmetry from~\eqref{e:SMPb} to~\eqref{e:SMPf}.
Spacetime supersymmetry is broken by interaction terms, explicitly omitted from~\eqref{e:SMPv}, \eqref{e:SMPb} and~\eqref{e:SMPf}, such as in the ``axilaton'' system~\cite{rBHM7}, and not at all unlike the Polonyi mechanism~\cite{Polonyi:1977pj}; for a recent discussion, see also~\cite{Ketov:2018uel}.
 
In addition, pseudo-scalars such as axions can be viewed as
boost generators (at least for constant profiles) between the observed and dual spacetimes~\cite{Freidel:2013zga}. First, note 
that the constant Kalb-Ramond field can be absorbed into a non-trivial symplectic form
(on its diagonal) after an $O(d,d)$ rotation~\cite{Freidel:2013zga}.
Thus the Kalb-Ramond two-form enters into an explicit non-commutativity of the modular spacetime, and
it can be used to rotate between observed and dual spacetime coordinates (as an explicit illustration of
relative, or observer-dependent, locality)~\cite{Freidel:2013zga}.
Since the Kalb-Ramond two-form dualizes into a pseudo-scalar in 4d, the 4d axion has the same features.
More generally, non-constant Kalb-Ramond profiles imply non-associative structure~\cite{Freidel:2013zga}.
Thus, axions are indicators of non-commutative (when constant) and non-associative (when propagating) structures in modular spacetime, respectively.

Finally, we note that the peculiar correlation between the visible and dark sectors, discussed for scalar, pseudo-scalar, fermionic
and vector degrees of freedom, can be also found in the gravitational and dual gravitational sectors.
Thus the observed gravity and dark energy are correlated via the scale of non-commutativity.
This might have interesting observable effects for the so-called $H_0$ tension~\cite{Verde:2019ivm}; see Ref.~\cite{mike}.

\section{Comments on phenomenological implications}
The most important general predictions of the chiral string worldsheet theory~\cite{Freidel:2013zga}
are: (1) the geometry of the dual spacetime determines the dark energy sector~\cite{Berglund:2019ctg}, and (2) the dual
matter degrees of freedom naturally appear as dark matter candidates, as discussed in the preceding section.
We note that, quite explicitly, the dark matter sector provides ``sources'' for the visible matter
sector. This follows from the coupling $ \frac{\l^8}{L^{10}}\brb{\phi}{\tilde{\phi}}$, 
as predicted by the doubled/non-commutative
set-up, and provides an explicit correlation between the dark matter sector and a visible sector.
Given the seesaw formula for the dark energy which relates the dark energy scale to the fundamental
length, which could be taken to be the Planck energy scale, then the dark matter is also sensitive to the dark energy.
So, the visible matter, dark matter and dark energy are all related.
This is consistent with the observational evidence presented in~\cite{Ho:2010ca},
as we have 
already alluded to in this article.

We emphasize that in our discussion of the hierarchy problem
the UV and IR scales are radiatively stable, and so is their product, the Higgs scale.
This new view on the hierarchy problem goes beyond the usual tools of effective field theory
due to explicit presence of the widely separated UV and IR scales.
The usual suggested approaches to the hierarchy problem: technicolor, SUSY and extra dimensions are
all within the canonical effective field theory.
In the context of string theory, effective field theory (and the approach to the
hierarchy problem via a SUSY effective field theory) can be naturally found via
string compactifications, but in that case one is faced with the issue 
of supersymmetry breaking
(and the fundamental question of ``measures'' on the string 
landscape/swampland~\cite{Danielsson:2018ztv, Obied:2018sgi, Andriot:2019wrs}.)
We claim that these issues are transcended in the general, doubled and non-commutative formulation
of string theory with a fundamentally bosonic and non-commutative formulation, wherein spacetime and matter
(and supersymmetry at the Planck scale)
can be viewed as emergent phenomena.

Next we comment on the seesaw formula, which mixes UV and IR scales, and the neutrino sector.
Such a seesaw would involve the neutrino and its dual partner.
Quite generically, the dual sector acts as a source for the visible sector, and the overall effect is to make the visible sector essentially massive.
This immediately provide a curious --- and ubiquitous --- mixing ``mass'' effect, where the length-scale ratio $\l^9/L^{10}$ may well be phenomenologically relevant; see Sect.~6. For every fermion-dual fermion pair, there is a mass-matrix of the general form $\left[\begin{smallmatrix}m&\l^9/L^{10}\\\l^9/L^{10}&\widetilde{m}\end{smallmatrix}\right]$, which induces well-known seesaw relations. This includes all Dirac mass-terms, as well as a mix of Dirac and Majorana terms, and may well provide a new way of generating neutrino masses.

Next, we comment on other phenomenological issues associated with the dual Standard Model.
In the dual QED sector, we should find a dual photon that is correlated to the visible photon, and
that is distinct from the dark photon of effective field theory. The correlation is proportional to
the fundamental length, and is finite even in the limit of zero momenta (deep infrared).
Also, the usual visible photon/dark photon coupling is subdominant to this term that
is inherent in our story. 
Similarly, in the dual of the weak sector of the Standard Model, we have a dual of the visible $Z$.
This dual of $Z$ should be distinguished from the usual $Z'$ by its sensitivity to the fundamental
length and by its correlation to the visible $Z$. 
These type of correlations might be found in correlated events in the accelerators, but which are
not products of any standard particle decays.

Finally, in the dual QCD we should find interesting phenomenology in the deep infrared, even though that
is a very difficult region to study in QCD. 
In particular, given the new view of the axion field in the
above discussion we have a possible new viewpoint of the strong CP problem in QCD.
The first observation here is that according to~\cite{Freidel:2013zga}, 
the constant Kalb-Ramond field mixes $x$ and $\tilde x$ spacetimes (it acts as a boost that linearly combines the spacetime and its dual in the context of a larger doubled and non commutative quantum spacetime). Also the commutator of dual spacetime coordinates is given by the constant $B$ Kalb-Ramond 2 form. So, for $B=0$ we get just the observed (4d) spacetime. Also its $H=\rd B$ field strength is trivially zero. But $H$ is dual (in 4d) to 
the axion ($a$), which is also constant. But $B$ is zero (there is no preferred background direction) and so this constant axion may be interpreted as a uniform distribution for the axion (whose constant values can be positive and negative). 
Now, focus on the QCD axion, relevant to the strong CP problem, and appearing in the CP violating term $a F\wedge F$. Averaging this term, linear in the axion, over a uniform distribution for this axion produces zero: ($\int_{-k}^k\rd a~ a=0$, with $k \to \infty$).
For a complete argument, we would have to study small fluctuations of the axion field in order to understand the robustness of this new viewpoint on the strong CP problem.

\section{Dark energy seesaw and the hierarchy problem}
\label{s:BHM}
The preceding discussion about dark energy, dark matter and the hierarchy problem is based on the generic non-commutative
formulation of string theory. We now present a more conventional realization of the above
analysis 
within a class of a specific 
discretuum of toy models~\cite{rBHM1,rBHM4, rBHM5, rBHM6, rBHM7}
that aim to realize de~Sitter space in string theory. In particular
several of the features of the above non-commutatively generalized phase-space reformulation of string theory are naturally captured due to the essential stringy nature of the models.

This family of models is constructed by starting with an
  F-theoretic~\cite{rFTh} type-IIB string theory spacetime, $W^{3,1} \times Y^4\times Y_\bot^2(\times T^2)$, where the complex structure of the zero-size ``hidden'' $T^2$ fiber of F-theory is identified with the axion-dilaton $\t\define\a\,{+}\,ie^{-\F}$ modulus.
Specifically, we compactify on $Y^4=\text{K3}$ or $T^4$ and let 
 the observable spacetime $W^{3,1}$ (via warped metric) vary over $Y_\bot^2$,  and $Y_\bot^2\to S^1 \times Z$, with the polar parametrization $\ell e^{z+i\q}=re^{i\q}$, where $z\define\log(r/\ell) \in Z$,
while $Y^4$ preserves supersymmetry. Finally, we deform $\t$ to vary {\em\/non-holomorphically,} only over $S^1\,{\subset}\,Y^2$.
By cross-patching two distinct solutions and by deforming the apparently singular metric into de~Sitter space, we get the final non-supersymmetric solution.
The codimension-2 solution $W^{3,1}\rtimes(S^1\times Z)$, has a positive cosmological constant, $\L$, along $W^{3,1}$, and the warped metric is~\cite{Cohen:1999ia}
\begin{equation}
 \rd s^2 = A^2(z)\, \bar g_{ab}\,\rd x^a \rd x^b - \ell^2 B^2(z)\,(\rd z^2 + \rd\q^2),
\end{equation}
where $\bar g_{ab}\,\rd x^a \rd x^b = \rd x_0^2 - e^{2\sqrt{\L}\,x_0}\,(\rd x_1^2 +\rd x_{2}^2 + \rd x_{3}^2) $ is the metric on $W^{3,1}$.
The two explicit solutions for $\t$ are~\cite{rBHM1}
\begin{alignat}9
 \t_I(\q) &{=}b_0+i\,g_s^{-1}\,e^{\w(\q-\q_0)}, \quad\text{and} \label{e:tau1}\\
\t_{I\!I}(\q) 
&{=}\big(b_0\pm g_s^{-1}\tanh[\w(\q{-}\q_0)]\big) \pm i\,\frac{g_s^{-1}}{\cosh[\w(\q{-}\q_0)]}. \label{e:tau2}
\end{alignat}
Given the stringy SL$(2;\mathbb{Z})$ monodromy of the axion-dilaton system over a transversal 2-plane $Y_\bot^2$ in the spacetime, these toy models exhibit S-duality.
 In generalizations where various moduli fields replace the axion-dilaton system,
this directly implies T-duality, 
which is covariantly realized in the generic phase-space approach~\eqref{e:MSA}--\eqref{e:CnCR1}.

We emphasize that these models are
a deformation of the stringy cosmic string (D7-brane in IIB string theory)~\cite{Greene:1989ya}, 
and as such represent effective
stringy solutions (in the sense of~\cite{Polchinski:1991ax}) and not just IIB supergravity solutions. That is, our solutions are
indeed found as deformations of certain classic F-theory backgrounds, but as codimension-2 solutions
they can be viewed as effective stringy solutions with an effective ``worldsheet''
description that is, to lowest order,  
doubled and generically non-commutative (as described by
equation~\eqref{e:MSA}). Thus our deformed stringy cosmic string solutions
are naturally equipped with a generalized geometric (and
non-commutatively doubled) spacetime structure, which to lowest order of the
doubled target space description directly connects to~\cite{Tseytlin:1990nb}.
Therefore certain generic features of this doubled description, such as the intensive effective
action, directly translate into certain geometric features of our models, discussed below.

In these string models 
the cosmological constant within the codimension-2 brane-world is determined by the anisotropy $\Delta \omega$ of the axion-dilaton system 
whose effective energy momentum tensor is given via
\begin{equation}
\!\!\!
 T_{\mu\nu}{-}{\textstyle\frac12}g_{\mu\nu}\,g^{\rho\s}T_{\rho\s}
 ={\cal G}_{\t\bar{\t}}\,\vd_{\mu}\t\vd_{\nu}\bar\t
 =\hbox{diag}[0,\cdots\!,\,0,\,\inv4\w^2\ell^{-2}],
\end{equation}
with ${\cal G}_{\t \bar{\t}}{=}\,{-}1/(\t{-}\bar\t)^2$, and
where $\ell$ is  the characteristic length-scale in the transversal 2-plane $Y_\bot^2$~\cite{rBHM5,rBHM6,rBHM7}:
\be
  \Lambda \sim \frac {{\Delta \omega}^2}{\ell^2}~~\text{implies}~~
  M_\L\,{\sim}\,M^2/M_P,
\label{e:seesaw}
\ee
relating the mass scales of the vacuum energy/cosmological constant ($M_\L$), particle physics, i.e., Standard Model ($M$), and the Planck scale ($M_P$). This {\it seesaw} formula can be seen to arise in two ways:
First, the formula~\eqref{e:seesaw}  may be understood as a consequence of dimensional transmutation, whereby the (modified) logarithmic nature of the transversal Green's function~\cite{rBHM1} (characteristic only of codimension-2 solutions) relates the length-scales $\ell$ and $\sqrt{\Lambda}$~\cite{rBHM5}.
Alternatively, the seesaw formula~\eqref{e:seesaw} follows from adapting Tseytlin's result for $\bar{S}$ to the models of~\cite{rBHM5,rBHM6,rBHM7}: In the denominator of 
the above formula,
the volume of the transversal 2-plane produces the length scale $\int_{Y_\bot^2}\!\sqrt{-g(x)}\,{\propto}\,\ell^2$; the numerator 
(with ${\Delta \omega}^2 \define \big(\omega^2 - \omega^2_c\,A^2{(z\,{=}\,0)}\big) $)
\be
  \int_{Y_\bot^2}\! \sqrt{-g(x)} 
  \big(R(x) +L_m\big) \propto {\Delta \omega}^2,
\ee
reproduces the anisotropy variance of these axion-dilaton profiles, whereas the remaining volume-integration renormalizes the Newton constant as required in~\cite{rBHM1,rBHM5}.
The anisotropy $\omega$ determines the above axion-dilaton stress tensor for the de~Sitter solution, and asymptotes to the Minkowski
cosmic brane limit $\omega_{c}$ at $z \to 0$. 
Note that in the F-theory limit, 
$\omega \to 0$ and $\omega_c \to 0$.
This singular supersymmetric configuration is deformed into a de~Sitter background by turning on an anisotropic axion-dilaton profile~\eqref{e:tau1}--\eqref{e:tau2}.
Thus $\Lambda$ that figures in the seesaw formula can be understood as being related to the cosmological breaking of supersymmetry.  
We stress that our discussion gives an argument for the existence of
de~Sitter background in string theory, albeit in its generic generalized-geometric and intrinsically
non-commutative formulation, which from the effective spacetime description is described
by our stringy models. One of the features of this doubled and generalized geometric 
description is that the effective spacetime action is intensive (as opposed to extensive), 
which directly translates into the seesaw formula for the cosmological constant~\eqref{e:seesaw}.

Note that more explicitly
\begin{equation}
\L_{D-2}\sim \frac {{\Delta \omega}^2}{\ell^2}
 =\Big(\frac{\pi}{D{-}3}\Big)^2
M_{D-2}^{~D-2}~(\ell\,M_{D-2})^2~\Big(\frac{M_D}{M_{D-2}}\Big)^{2D-4}~.
   \label{e:energydensity}
\end{equation}
 In our primary case of interest, of a minimal (simple) supersymmetry-preserving compactification to $D=6$ dimensions, this becomes
\begin{equation}
  \L\sim \frac {{\Delta \w}^2}{\ell^2}
  =\frac{\pi^2}9~\ell^2~\frac{M_6^{~8}}{M_4^{~2}},
  \quad\xrightarrow[M_6\,\mapsto\,M]{1/\ell\,\sim\,M_4\,\mapsto\,M_P}\quad
  M_\L\,{\sim}\,M^2/M_P,
\label{e:seesawD6}
\end{equation}
and relates the mass scales of the vacuum energy/cosmological constant ($M_\L$), particle physics/Standard Model ($M_6\mapsto M$), and the Planck scale ($M_P$). 

More precisely, the $D$- and the $(D{-}2)$-dimensional characteristic (Planck) mass-scales are related by the exact expression~\cite{rBHM1,rBHM4,rBHM7}\footnote{Note: the exponential factor~\cite[Eq.~(15)]{rBHM4} was inadvertently omitted in Ref.~\cite[Eq.~(3.3)]{rBHM7}. Also, note that $0\leqslant\Gamma(x;y)\define\big(\Gamma(x)-\g(x;y)\big)\leqslant\Gamma(x)$, and both incomplete gamma functions range from 0 to $\Gamma(x)$.}
\begin{equation}
 M_{D-2}^{D-4}
 = M_D^{D-2}\,2\p\ell^2\,|z_0|^{-\frac{D-1}{2(D-2)}}\,e^{z_0}\,
   \Gamma_\pm\Big(\frc{D-3}{2(D-2)};\frc1{|z_0|}\Big)
\label{e:normalization}
\end{equation}
where $z_0$ is the radial distance (in units of $\ell$) from the $(D{-}2)$-dimensional brane-World to the boundary of $\mathscr{Y}_\bot^2$, and $\Gamma_\pm$ denotes the ``$[0,\frc1{|z_0|}]$-incomplete gamma function'' for $z_0<0$ and its complement for $z>0$. For $D=6$, this yields
\begin{alignat}9
  &M_4^{~2} \,{=}\, \z_0\,|z_0|^{-\frac58}\,e^{z_0}~M_6^{~4}\,\ell^2,~~
    M_4 \,{=}\, \sqrt{\z_0}\,|z_0|^{-\frac5{16}}\,e^{z_0/2}~\frac{M_6^{~2}}{M_\ell},
 \label{e:expHM4}\\
  &\text{where}~~
   0\leqslant\z_0\define2\p\Gamma\big(\frc38;\frc1{|z_0|}\big)
    \leqslant2\p\Gamma\big(\frc38\big)\approx14.89,
\end{alignat}
focusing on the $z_0>0$ case of~\eqref{e:normalization}, since that enables the {\em\/exponential hierarchy\/} $M_4\gg M_6$.\footnote{In the ``naked singularity to brane-World'' coalescing limit $\lim_{z_0\to0}\Gamma(\frc38;\frc1{|z_0|})|z_0|^{-5/8}e^{z_0}=0$ since $\Gamma(\frc38;\frc1{|z_0|})$ vanishes faster than any (negative) power can diverge. In turn, moving the naked singularity away 
from the brane-World by keeping $z_0\neq0$ makes the hierarchy grow exponentially,~$\sim e^{z_0}$.}
In turn, solving~\eqref{e:expHM4} for $M_6$, the cosmological constant~\eqref{e:seesawD6} becomes:
\begin{equation}
  \L\sim\frac{\pi^2}{9\z_0}|z_0|^{5/4}e^{-2z_0}~\frac{M_4^{~2}}{\ell^2}
        =\frac{\pi^2}{9\z_0}|z_0|^{5/4}e^{-2z_0}~M_4^{~2}M_\ell^{~2},
 \label{e:expHL}
\end{equation}
which reveals the effect of the $z_0$-driven {\em\/exponential hierarchy\/} in the ``axilaton'' models~\cite{rBHM1,rBHM4,rBHM5,rBHM6,rBHM7}. Therefore, the {\em\/primordial\/} ``size-of-the-universe-scale'' $L$ in~\eqref{e:SMPv}--\eqref{e:SMPf} is free to be naturally within one or two orders of magnitude of the stringy fundamental length-scale $\l$, resulting in a phenomenologically realistic scale introduced by the mixing terms~\eqref{e:SMPv}--\eqref{e:SMPf}!

The seesaw expression appears to be technically natural.
That is, when $M_P \to \infty$ the cosmological constant
scale goes to zero, and in that case the dual space curvature is zero, and we get a flat dual space, and thus
enhanced symmetry. This is precisely what 't Hooft naturalness asks of us: when the physical cutoff 
in some theory goes to infinity,
the small parameters in the theory vanish and should be protected by some hidden symmetry.
It is tempting to relate that hidden symmetry to supersymmetry.
However, this appears to be a bit too naive. Such conjectural supersymmetry restoration requires the vanishing of the Ricci tensor, which requires $\w\to0$: the vanishing $\L\propto\Delta\w^2\to0$ is necessary, but not sufficient. Letting $\L\to0$ by sending $M_4\to\infty$ can be forced by letting $z_0\to\infty$ in~\eqref{e:expHM4}, the geometrical meaning of which is that $\mathscr{Y}^2_\bot\to S^1$ ``at infinity,'' ---indicating
some singular dimensional collapse.

This seesaw formula can be rewritten in a form that
is even more appealing by relating the scale $M$ to the Planck scale via an exponential factor
$M = M_p \exp (- \const M_p/M_i) $
and thus the vacuum energy density $M_p^2 \Lambda$ is given as
\be
 M_\L^4 = M_p^4 \exp\{- 8\,\const M_p/M_i\}
\ee
where $M_i$ correspond to what is roughly an effective scale that numerically corresponds to the inflation scale 
(see also~\cite{Obied:2018sgi}).
Note that~\eqref{e:expHL} comes close to this, except 
that it is hard to set
$z_0\to M_4/M_i$ except by hand.
Nevertheless, our toy model, even though realized in the conventional spacetime interpretation of string theory,
does illustrate the main features of the generic non-commutative and doubled formulation of string theory,
at least when it comes to the dark energy seesaw and its relation to the separation of scales associated
with the hierarchy problem.
Finally, we comment on the UV and IR scales in the section on dark matter and string theory. 
the UV (or the non-commutativity scale) should be considered as an effective scale to be empirically determined. 
 In turn, $L$ is the primordial IR cutoff.

\section{A non-perturbative formulation of string theory}
\label{s:non-pert}

In view of the preceding discussion regarding the generic formulation of string theory and its relation to the problem of
dark energy and dark matter, we now propose a non-perturbative formulation of string theory (and its M- and F-theory avatars). Indeed, the chiral string worldsheet theory
offers such a new view on the fundamental question of a non-perturbative formulation of quantum gravity~\cite{Freidel:2013zga}
by noting the following:
in the chiral string worldsheet
description the target space is found to be a modular space (quantum spacetime), but 
the same can be also said of the worldsheet.
If the string worldsheet is made modular in its chiral 
formulation, by doubling of $\tau$ and $\sigma$, so
that  $\X (\tau, \sigma)\to\widehat{\X} (\tau, \sigma)$ can  be
in general viewed as an infinite dimensional matrix (acting on the basis of Fourier components of $\tt$ and $\ts$, the
doubles of $\tau$ and $\sigma$, respectively), then the corresponding chiral string worldsheet action becomes
\be
\int_{\t,\s}\! \Tr \big[ \partial_{\tau} \widehat{\X}^A \partial_{\sigma} \widehat{\X}^B (\omega_{AB} + \eta_{AB}) - 
\partial_{\sigma} \widehat{\X}^A H_{AB} \partial_{\sigma} \widehat{\X}^B \big] ,
\ee
where the trace is over the (suppressed) matrix indices.
The matrix elements then emerge as the natural partonic degrees of freedom.
We arrive at a {\it non-perturbative quantum gravity} by replacing the $\s$-derivative with a
commutator involving one extra $\widehat{\X}^{26}$ (with $A=0,1,2,\dots,25$)\footnote{That the canonical worldsheet of string theory might become non-commutative in a deeper, non-perturbative formulation, was suggested in~\cite{Atick:1988si}.}:
\be
\partial_{\sigma} \widehat{\X}^A \to [\widehat{\X}^{26}, \widehat{\X}^A] .
\ee
This dictionary suggests the following fully interactive and non-perturbative formulation of the chiral string worldsheet theory
in terms of a (M-theory-like) matrix model form of the above chiral string 
action (with $a,b,c=0,1,2,\dots,25, 26$ )
\be
\int_\t\! \Tr \big[ \partial_{\tau} \widehat{\X}^a [\widehat{\X}^b, \widehat{\X}^c] \eta_{abc}  - 
H_{ac} [\widehat{\X}^a, \widehat{\X}^b] [\widehat{\X}^c, \widehat{\X}^d] H_{bd}\big],
\ee
where the first term is of a Chern-Simons form and the second of the Yang-Mills form, and $\eta_{abc}$ contains both 
$\omega_{AB}$ and $\eta_{AB}$.
This is then the non-perturbative ``gravitization of the quantum''~\cite{Freidel:2013zga}.
We remark that in this non-perturbative matrix theory-like formulation of the chiral string (and quantum gravity),
the matrices emerge from the modular worldsheet, and the fundamental commutator from the
Poisson bracket with respect to the dual world sheet coordinates (of the modular/quantum world sheet) --- that is,
quantum gravity ``quantizes'' itself, and thus quantum mechanics originates in quantum gravity.
(However, this formulation should be distinguished from Penrose's ``gravitization of the quantum'' and gravity-induced ``collapse of the wave function''~\cite{Penrose:2014nha}.
Also note some similarity of the chiral string worldsheet 
formulation, in its intrinsic non-commutative form, to
the most recent proposal by Penrose regarding ``palatial'' twistor theory 
\cite{Penrose:2015lla}.)

At this point we also recall that the authors of~\cite{Atick:1988si} explicitly state in the conclusion of their paper:
``(1) The density of gauge invariant degrees of freedom, per unit energy, per unit spacetime volume, is much less in the
proper formulation of string field theory than in any ordinary relativistic field theory\dots\
(2) The translation degree of freedom of the string center of mass is in some sense doubled\dots\
(3) The familiar continuous world sheets should be replaced, in the proper formulation of the classical theory,
by some less continuous structure, perhaps related to continuum world sheets the way quantum phase space is related to classical phase space."
This prescient prediction is remarkably close to our non-perturbative formulation!

In thinking about non-perturbative matrix model formulations of string theory it is natural to invoke the IIB matrix model~\cite{Ishibashi:1996xs}, based on D-instantons as well as the matrix model of 
M-theory~\cite{Banks:1996vh}, based on D0-branes. However, these matrix models lack very important covariant
properties associated with F-theory and M-theory. In our proposal we can do better.
Given our new viewpoint we can suggest a 
new covariant 
non-commutative matrix model formulation of F-theory, by also writing in the large $N$ limit 
$\pa_{\tau}\widehat{\X}^C = [\widehat{\X}, \widehat{\X}^C]$, 
in terms of commutators of two (one for $\pa_\s\widehat{\X}^C$ and one for $\pa_{\tau}\widehat{\X}^C$) extra $N\,{\times}\,N$ matrix valued chiral $\widehat{\X}$'s.
Notice that, in general, we do not need an overall trace, and so the action can be viewed as a matrix, rendering the entire non-perturbative formulation of F-theory as
purely quantum in the sense of the original matrix formulation of quantum mechanics 
by Born-Jordan and Born-Heisenberg-Jordan~\cite{history} 
\begin{equation}
 \S_{\text{ncF}}\,{=}\,\frac{1}{4\pi}
[{\widehat{\X}}^{a},  {\widehat{\X}}^{b}] [{\widehat{\X}}^{c},  {\widehat{\X}}^{d}]  f_{abcd},
\end{equation}
where instead of $26$ bosonic $\widehat{\X}$ matrices one would have $28$, with supersymmetry emerging in 10($+$2) dimensions from this underlying
bosonic formulation. This formulation realizes the $SL(2,\ZZ)$ symmetry of IIB string theory.
In this non-commutative matrix model formulation of F-theory, in general,
$f_{abcs} ({\widehat{\X}})$ is a dynamic background, determined by the matrix analog of the vanishing of
the relevant beta function.
By T-duality, the new covariant M-theory matrix model reads as 
\begin{equation}
 \S_{\text{ncM}}\,{=}\,\frac{1}{4\pi}  
\int_{\tau}\!
\big(\pa_{\tau} \widehat{\X}^i [{\widehat{\X}}^{j},  {\widehat{\X}}^{k}] g_{ijk} - [{\widehat{\X}}^{i},  {\widehat{\X}}^{j}][{\widehat{\X}}^{k},  {\widehat{\X}}^{l}] h_{ijkl}\big),
\end{equation}
with 27 bosonic $\widehat{\X}$ matrices, with supersymmetry emerging in 11 dimensions.
Once again, the backgrounds $g_{ijk} ({\widehat{\X}})$, and $h_{ijkl} ({\widehat{\X}})$ are fully dynamical, should be determined by a matrix-analog of the Renormalization Group (RG) equation, and the vanishing of the
corresponding beta function.

Note that in this approach
holography~\cite{tHooft:1993dmi} (such as AdS/CFT~\cite{Maldacena:1997re}, which can be viewed as a ``quantum Jarzynski equality on the space of geometrized RG flows''~\cite{Minic:2010pw}) is emergent in a particular semi-classical ``extensification''
of quantum spacetime, in which the dual spacetime degrees of freedom are also completely decoupled.
 The relevant information about
 $\w_{AB},\eta_{AB}$ and $H_{AB}$ is now contained in the new dynamical backgrounds $f_{abcd}$ in F-theory, and $g_{ijk}$ and $h_{ijkl} $ in M-theory.
This proposal offers a new formulation of covariant Matrix theory in the M-theory limit~\cite{Minic:1999js},
which is essentially a partonic formulation: Strings emerge from partonic constituents in a certain limit. This new matrix formulation is fundamentally bosonic and thus it is reminiscent of bosonic M-theory~\cite{Horowitz:2000gn}.
The relevant backgrounds $g_{ijk}$ and $h_{ijkl} $ should be determined by the matrix RG equations.
Also, there are lessons here for the new concept of ``gravitization of quantum theory'' as well as the idea that dynamical Hilbert spaces or 2-Hilbert spaces (here represented by matrices) are fundamentally needed
in quantum gravity~\cite{twohilbert}.
This matrix like formulation should be understood as a general non-perturbative formulation of 
string theory. In this partonic (quantum spacetime) formulation closed strings (as well as branes) are collective excitations, in turn constructed from the product of open string fields. Similarly, our toy model can be understood as a collective excitation in this more
fundamental ``partonic'' formulation. The observed classical spacetime emerges as an 
``extensification''~\cite{Freidel:2013zga}, 
in a particular limit, out of the basic building blocks of quantum spacetime.
Their remnants can be found in the low energy bi-local quantum fields, with bi-local 
quanta, which
were a motivation for our discussion of dark matter in string theory.

Finally, it is an old realization that the 10d superstring can be found as a solution of the bosonic string theory 
\cite{Casher:1985ra}.
The authors~\cite{Casher:1985ra} explicitly state in their abstract that ``consistent closed ten-dimensional superstrings, i.e., the two $N\,{=}\,1$ heterotic strings and the two $N\,{=}\,2$ superstrings, are contained in the 26-dimensional bosonic closed string theory. The latter thus appears as the fundamental string theory.''
This is precisely what we have in our proposed non-perturbative formulation.
(Such a bosonic formulation is also endowed with higher mathematical symmetries, as already observed
in~\cite{Moore:1993zc}.)
Supersymmetry (in the guise of M- and F-theory) is emergent from our non-perturbative and seemingly entirely
bosonic formulation in a similar fashion.
This should allow going around the obvious problems raised by the apparent falsification of supersymmetry at
the observable LHC energies.

\section{Concluding remarks}
In this paper we have related
 the problems of dark energy and dark matter
 with the hierarchy problem,
 in the context of a general non-commutative formulation of string theory,
 wherein dark energy is generated by the dynamical geometry of dual
spacetime.
 In particular, dark matter stems from the degrees of freedom dual to the visible matter.
This generic formulation of string theory is sensitive to both the IR and UV scales. The Standard Model (Higgs) scale is radiatively stable by being a geometric 
mean of these two radiatively stable scales, which clearly goes beyond the reach of effective field theory.
 We also have commented on various phenomenological signatures of this new approach to dark energy, dark matter and the hierarchy problem in the context of string theory, and
the realization of this new view on the
hierarchy problem within a discretuum of toy models based on a non-holomorphic deformation of stringy cosmic strings.
Finally, we have presented a proposal for a new
non-perturbative formulation of string theory, which sheds light on both M- and F-theory, and illuminates
issues related to supersymmetry and holography.

In conclusion, we point out that the sequester mechanism discussed in this paper can be used to stabilize the moduli,
and that it offers a new view on SUSY breaking beyond effective field theory, based on T-duality and
intrinsic non-commutativity of string theory.
In our approach SUSY might be important for fixing the zero value of the cosmological constant only in the limit of
infinite Planck scale in 4d, and, also, for stability of locally emergent Minkowski spacetime.
Also, in our approach, the Standard Model does not have to be realized via the Kaluza-Klein mechanism and string compactifications,
but in the context of ``extensification'' of the non-pertubative formulation of chiral string worldsheet 
theory, in
which case one should look for robust non-commutative structures in the Standard Model, as indicated
by Connes approach to the so-called non-commutative geometry of the Standard Model~\cite{connes} and its
phenomenology~\cite{Chamseddine:2019fjq,Devastato:2019grb, Aydemir:2013zua}.
This new view of string theory should bring about a different viewpoint on the vacuum selection problem
(along the lines of the attractor solutions found in~\cite{Argyriadis:2019fwb}), 
as well as the selection of a robust quantum matter
sector which is mutually consistent with the quantum gravitational sector and their respective duals.

\section*{Acknowledgements}
We would like to thank J.A.~Argyriadis, D.~Edmonds,  L.~Freidel, V.~Jejjala, M.~Kavic, J.~Kowalski-Glikman, 
Y.-H.~He, R.~Leigh and T.~Takeuchi for discussions.
PB would like to thank 
the Simons Center for Geometry
and Physics and the CERN Theory Group, for their hospitality,
and TH is grateful to 
the Department of Physics, University of Maryland, 
and the Physics Department of 
the University of Novi Sad, Serbia, for hospitality and resources. 
The work of DM is supported in part by Department of Energy 
(under DOE grant number DE-SC0020262)
and the Julian Schwinger Foundation. D.M. is grateful to Perimeter Institute for hospitality and support.

\bibliographystyle{unsrt}

\end{document}